\documentclass[conference]{IEEEtran}

\usepackage{cite}
\usepackage{amsmath,amssymb,amsfonts}
\usepackage{url}
\usepackage{booktabs}
\usepackage{savefnmark}
\usepackage{xfrac}

\usepackage[caption=false,font=footnotesize,labelformat=simple]{subfig}

\usepackage{forest}
\tikzset{keep/.style = {rectangle, draw, align=center, anchor=north, font=\tiny, inner sep=1.5pt}}
\tikzset{edge from parent path={(\tikzparentnode.south) -- (\tikzchildnode.north)}}
\tikzset{level distance = 3.5ex}
\newcommand{\vmax}{\vphantom{abcdefghijklmnopqrstuvwxyzABCDEFGHIJKLMNOPQRSTUVWXYZ[]}}
\newcommand{\token}[1]{\texttt{\vmax{}'#1'}}

\usepackage{listings}
\lstdefinelanguage{pseudo}{
  morekeywords={procedure,begin,end,return,forall,in,do,while,if,then,elif,else,continue,break,parallel,and,not,global,assert},
  sensitive=true,
  morecomment=[s]{(*}{*)},
  morestring=[b][\ttfamily]',
  columns=fullflexible,
  tabsize=2,
  basicstyle={\rmfamily},
  numbers=left,
  xleftmargin=.75cm,
  mathescape,
  frame=tb,
  numberblanklines=false,
}
\lstdefinelanguage{clike}{
  sensitive=true,
  morecomment=[s]{/*}{*/},
  morecomment=[l]{//},
  morestring=[b][\ttfamily]',
  morestring=[b][\ttfamily]",
  tabsize=2,
  basicstyle={\ttfamily\footnotesize},
  showstringspaces=false,
}
\lstdefinelanguage{javalike}{
  sensitive=true,
  morecomment=[s]{/*}{*/},
  morecomment=[l]{//},
  morestring=[b][\ttfamily]',
  morestring=[b][\ttfamily]",
  tabsize=2,
  basicstyle={\ttfamily\footnotesize},
  showstringspaces=false,
  numbers=left,
  frame=tb,
  numberblanklines=false,
  breaklines=true,
  postbreak=\mbox{\textcolor{red}{$\hookrightarrow$}\space},
}

\usepackage{pifont}
\newcommand{\pass}{\text{\ding{51}}}
\newcommand{\fail}{\text{\ding{55}}}
\newcommand{\unres}{\text{\textbf{?}}}

\newcommand{\alg}[1]{\textsc{#1}}
\newcommand{\DDMIN}{\alg{ddmin}}
\newcommand{\HDD}{\alg{hdd}}
\newcommand{\TMIN}{\alg{tmin}}
\newcommand{\HOIST}{\alg{hoist}}
\newcommand{\HDDH}{\alg{hddh}}
\newcommand{\TAGNODES}{\alg{tagNodes}}
\newcommand{\PRUNE}{\alg{prune}}
\newcommand{\TRANSFORM}{\alg{transform}}

\begin{document}

\title{Extending Hierarchical Delta Debugging with Hoisting}

\author{
  \IEEEauthorblockN{D\'aniel Vince, Ren\'ata Hodov\'an, Daniella B\'arsony and \'Akos Kiss}
  \IEEEauthorblockA{
    Department of Software Engineering \\
    University of Szeged \\
    Szeged, Hungary \\
    Email: \{vinced,hodovan,bella,akiss\}@inf.u-szeged.hu
  }
}

\maketitle

\begin{abstract}
Minimizing failing test cases is an important pre-processing step on the path of debugging.
If much of a test case that triggered a bug does not contribute to the actual failure, then the time required to fix the bug can increase considerably.
However, test case reduction itself can be a time consuming task, especially if done manually.
Therefore, automated minimization techniques have been proposed, the minimizing Delta Debugging (DDMIN) and the Hierarchical Delta Debugging (HDD) algorithms being the most well known.
DDMIN does not need any information about the structure of the test case, thus it works for any kind of input.
If the structure is known, however, it can be utilized to create smaller test cases faster.
This is exemplified by HDD, which works on tree-structured inputs, pruning subtrees at each level of the tree with the help of DDMIN.

In this paper, we propose to extend HDD with a reduction method that does not prune subtrees, but replaces them with compatible subtrees further down the hierarchy, called hoisting.
We have evaluated various combinations of pruning and hoisting on multiple test suites and found that hoisting can help to further reduce the size of test cases by as much as 80\% compared to the baseline HDD.
We have also compared our results to other state-of-the-art test case reduction algorithms and found that HDD extended with hoisting can produce smaller output in most of the cases.
\end{abstract}

\begin{IEEEkeywords}
test case minimization, hierarchical delta debugging, hoisting
\end{IEEEkeywords}

\section{Introduction}
\label{sec:into}

Our software can fail, so our software will fail.
If we are lucky enough, we have a record of the events or inputs that triggered the failure.
If we are even luckier, the failure is reproducible.
In this case, a lucky engineer will get the task of fixing the problem.

Usually, the observed problem is only a symptom, and the root of it has to be found first in order to get the bug fixed.
The record of the events or inputs that triggered the failure -- i.e., the test case -- can help here.
However, this test case is often a mixture of relevant and irrelevant information.
If much of the test case is irrelevant, i.e., it does not contribute to the failure, then the engineering time and effort required to fix the bug can increase considerably.
Therefore, the minimization of failure-inducing test cases is an important first step on the path of debugging.
However, it is of limited benefit if engineer-hours spent on bug fixing are simply converted to engineer-hours spent on manual test case reduction.

One of the most well-known techniques to automate reduction is the minimizing Delta Debugging algorithm (\DDMIN) by Zeller and Hildebrandt~\cite{zeller-esecfse-1999,hildebrandt-issta-2000,zeller-tse-2002}, working on all kinds of test cases without the need for any information about their structure.
It has been realized, however, that if the structure of the test cases is known, that knowledge can be utilized to create smaller results faster.
Misherghi and Su have introduced the Hierarchical Delta Debugging (\HDD) algorithm~\cite{misherghi2006hdd,misherghi2007hdd}, built on \DDMIN, that works on tree-structured inputs (e.g., on any input format that has a context-free grammar) and prunes unnecessary subtrees of the test case during reduction.
These foundational works have inspired many follow-up research:
some papers focused on improving their performance~\cite{hodovan2016practical,hodovan2017tree,hodovan2017coarse,kiss2018hddr}, while others aimed for smaller results~\cite{hodovan2016modernizing}.

In this paper, we also focus on the size aspect of the reduced test cases.
We propose to extend the pruning-based reduction approach of \HDD, where subtrees of the test case are removed, with a technique called hoisting, where subtrees are replaced with compatible subtrees further down the hierarchy.
Therefore, we define the algorithmic framework of hoisting and describe its potential combinations with pruning.
We have evaluated the introduced approaches and found that hoisting can help to further reduce the size of test cases by as much as 80\% compared to the baseline \HDD.

The rest of the paper is organized as follows:
first, in Section~\ref{sec:background}, we give a brief overview of {\DDMIN} and \HDD, to make this paper self-contained.
Then, in Section~\ref{sec:hoisting}, we show some examples where pruning-based reduction can be improved upon, and describe and formalize the idea of hoisting.
In Section~\ref{sec:results}, we evaluate the effects of hoisting with the help of a prototype implementation, and we present our experimental results.
In Section~\ref{sec:relatedwork} we discuss related work, and
finally, in Section~\ref{sec:summary} we summarize our work and conclude the paper.

\section{Background}
\label{sec:background}

\begin{figure*}[th!]
\small
\makebox[\linewidth]{\hrulefill}
Let $test$ and $c_\fail$ be given such that $test(\emptyset) = \pass \wedge test(c_\fail) = \fail$ hold.\\
The goal is to find $c'_\fail = ddmin(c_\fail)$ such that $c'_\fail \subseteq c_\fail$, $test(c'_\fail) = \fail$, and $c'_\fail$ is 1-minimal.\\
The \textit{minimizing Delta Debugging algorithm} $ddmin(c)$ is
\begin{align*}
ddmin(c_\fail) &= ddmin_2(c_\fail, 2)  \text{ where} \\
ddmin_2(c'_\fail, n) &= \left\{ \begin{alignedat}{-1}
  & ddmin_2(\Delta_i, 2)                    && \text{\hphantom{else }if } \exists i \in \{1, \ldots, n\} \cdot test(\Delta_i) = \fail  \text{ (``reduce to subset'')} \\
  & ddmin_2(\nabla_i, \max(n - 1, 2))       && \text{else if } \exists i \in \{1, \ldots, n\} \cdot test(\nabla_i) = \fail  \text{ (``reduce to complement'')} \\
  & ddmin_2(c'_\fail, \min(|c'_\fail|, 2n)) && \text{else if } n < |c'_\fail|  \text{ (``increase granularity'')} \\
  & c'_\fail                                && \text{otherwise (``done'').}
\end{alignedat} \right.
\end{align*}
where $\nabla_i = c'_\fail - \Delta_i$, $c'_\fail = \Delta_1 \cup \Delta_2 \cup \ldots \cup \Delta_n$, all $\Delta_i$ are pairwise disjoint, and $\forall \Delta_i \cdot |\Delta_i| \approx |c'_\fail| / n$ holds.\\
The recursion invariant (and thus precondition) for $ddmin_2$ is $test(c'_\fail) = \fail \wedge n \leq |c'_\fail|$.\\
\makebox[\linewidth]{\hrulefill}
\caption{The Minimizing Delta Debugging Algorithm.}
\label{fig:ddmin}
\end{figure*}

The minimizing Delta Debugging (\DDMIN) algorithm~\cite{zeller-esecfse-1999,hildebrandt-issta-2000,zeller-tse-2002} is a systematic iterative approach for reducing a test case while keeping some interesting property invariant.
The algorithm works on a set of atomic units representing parts of the test case.
First, this set of units is split into two subsets of roughly equal size, and both subsets are investigated for whether they still have the interesting property of the original test case.
If the property is kept in any of the subsets, reduction was successful and a new iteration starts with the found subset, otherwise the granularity is refined by doubling the splitting.
The subsets of the new partitioning are investigated again, one by one, as well as their complements.
I.e., it is checked whether keeping or removing any of the subsets leads to an interesting smaller test case.
Again, if any of the investigated test cases keeps the property in question, it will be used as the input to the next iteration, otherwise the splitting is increased.
The iteration continues until the granularity reaches the unit level, when it is proven to have found a so-called 1-minimal result, a local minimum where the removal of any single unit from the test case causes the loss of the interesting property.

The algorithm has its roots in the isolation of failure-inducing code changes, which is visible in its terminology.
For the algorithm, a test case is composed of elementary changes or deltas, denoted as $\delta_i$, whence the algorithm got its name.
A set of changes is also called a configuration, usually denoted by $c$.
The outcome of a program run on a specific configuration is determined by a testing function, and it can be either \textsc{fail} (also written as \fail) if the test case induced the original failure, \textsc{pass} (also written as \pass) if the test succeeds, or \textsc{unresolved} (written as \unres) if the result is indeterminate.
The set of all changes, i.e., the initial configuration that triggers a failing run is denoted by $c_\fail$.
Although the algorithm is often applied to the simplification of program inputs where the term ``change'' is not an intuitive fit to the units of a test case (e.g., to characters or lines of a text file) and the algorithm also has use cases where the interesting property of a test case is not a program failure.
Most authors, including us, follow the original notation for historical reasons.
For the sake of completeness, Figure~\ref{fig:ddmin} gives Zeller and Hildebrandt's latest formulation of the minimizing Delta Debugging algorithm~\cite{zeller-tse-2002}.

If a test case that is to be reduced has some mandatory structure over its units, which is quite typical for inputs to a program, {\DDMIN} may work suboptimally.
The configuration partitioning during the iterations may be completely unaligned with the boundaries of the structural elements of the input, leading to incorrectly formatted, non-reproducing, and thus useless test cases.
The goal of the Hierarchical Delta Debugging (\HDD) algorithm~\cite{misherghi2006hdd} is to avoid such superfluous steps by not testing format-breaking configurations.
To achieve that goal, it works on hierarchical tree-structured input representations (e.g., on parse trees, abstract syntax trees, or XML DOM trees) and applies the minimizing Delta Debugging algorithm to the levels of a tree, progressing downwards from the root to the leaves.

The pseudocode formulation of {\HDD} as defined by Misherghi and Su~\cite{misherghi2006hdd} is shown in Figure~\ref{fig:hdd}.
In the algorithm, the auxiliary routine {\TAGNODES} collects the nodes at a given level of the tree, then {\DDMIN} is invoked on those nodes, and finally {\PRUNE} applies the result of Delta Debugging to the tree.
I.e., for \HDD, configurations are sets of tree nodes at a given level and removal of a node causes the removal of the whole subtree rooted at that node.
In a later variant of \HDD, ``pruning'' of a node has been reinterpreted as its replacement with the minimal applicable syntactically correct fragment to reduce the number of test attempts at incorrectly formatted configurations even further~\cite{misherghi2007hdd}.
If {\HDD} is iterated until a fixed-point is reached, denoted as \HDD$^*$, it gives a 1-tree-minimal result, i.e., if any single node is removed from the tree, the new test case will not be interesting anymore.

\begin{figure}[t]
\begin{lstlisting}[language=pseudo]
procedure $\HDD(\textit{input\_tree})$
    $\textit{level} \leftarrow 0$
    $\textit{nodes} \leftarrow \TAGNODES(\textit{input\_tree}, \textit{level})$
    while $\textit{nodes} \neq \emptyset$ do
        $\textit{minconfig} \leftarrow \DDMIN(\textit{nodes})$
        $\PRUNE(\textit{input\_tree}, \textit{level}, \textit{minconfig})$
        $\textit{level} \leftarrow \textit{level} + 1$
        $\textit{nodes} \leftarrow \TAGNODES(\textit{input\_tree}, \textit{level})$
    end while
end procedure
\end{lstlisting}
\caption{The Hierarchical Delta Debugging Algorithm.}
\label{fig:hdd}
\end{figure}

\section{Hoisting}
\label{sec:hoisting}

Although {\HDD} performs better on structured inputs than \DDMIN, there is still room for improvement.
Several improvements have already been proposed, often by preprocessing the tree representation {\HDD} is working on, e.g., by hiding some tokens from {\HDD} and {\DDMIN} to reduce the number of nodes that have to be considered, by collapsing (a.k.a. squeezing) multiple nodes into one for the same reason~\cite{hodovan2017tree}, or by rotating recursive structures of the tree to reduce its height~\cite{hodovan2017coarse}.
However, these transformations do not change the core structure of the tree, i.e., the test case generated (or, serialized) from the preprocessed tree will still be the same as the original input.
Because of this, and because of the way {\HDD} works, an \HDD-reduced test case -- although being 1-tree-minimal -- may contain structural elements that a human expert would remove.

\subsection{Examples}
\label{sec:hoisting-ex}

\newsavebox\hwcbox
\begin{lrbox}{\hwcbox}
\begin{lstlisting}[language=clike]
int main() {
  if (1) {
    printf("Hello world!\n");
  }
}
\end{lstlisting}
\end{lrbox}

\newsavebox\hwctreebox
\begin{lrbox}{\hwctreebox}
\begin{forest}
for tree={parent anchor=south, child anchor=north, l=2ex, s sep=.25ex, tier/.pgfmath=level()}
[\vmax *, keep
  [\vmax externalDeclaration, keep
    [\vmax functionDefinition, keep
            [\vmax typeSpecifier, keep 
              [\token{int}, keep] 
            ]
      [\vmax directDeclarator, keep
        [Identifier\\\token{main}, keep]
        [\vmax *, keep
          [\token{(}, keep] 
          [\token{)}, keep] 
        ]
      ]
      [\vmax compoundStatement, keep
        [\token{\{}, keep] 
            [\vmax selectionStatement, keep 
              [\token{if}, keep] 
              [\token{(}, keep] 
              [\vmax primaryExpression, keep
                [Constant\\\token{1}, keep]
              ]
              [\token{)}, keep] 
              [\vmax statement, keep
                [\vmax compoundStatement, keep
                  [\token{\{}, keep] 
                  [\vmax *, keep
                    [\vmax expressionStatement, keep
                        [\vmax postfixExpression, keep 
                          [Identifier\\\token{printf}, keep]
                          [\vmax *, keep
                            [\token{(}, keep] 
                                  [StringLiteral\\\token{"Hello world!\textbackslash{}n"}, keep] 
                            [\token{)}, keep] 
                          ]
                        ]
                      [\token{;}, keep] 
                    ]
                  ]
                  [\token{\}}, keep] 
                ]
              ]
            ]
        [\token{\}}, keep] 
      ]
    ]
  ]
]
\end{forest}
\end{lrbox}

\newsavebox\hwchoisttreebox
\begin{lrbox}{\hwchoisttreebox}
\begin{forest}
for tree={parent anchor=south, child anchor=north, l=2ex, s sep=.25ex, tier/.pgfmath=level()}
[\vmax *, keep
  [\vmax externalDeclaration, keep
    [\vmax functionDefinition, keep
            [\vmax typeSpecifier, keep 
              [\token{int}, keep] 
            ]
      [\vmax directDeclarator, keep
        [Identifier\\\token{main}, keep]
        [\vmax *, keep
          [\token{(}, keep] 
          [\token{)}, keep] 
        ]
      ]
                [\vmax compoundStatement, keep
                  [\token{\{}, keep] 
                  [\vmax *, keep
                    [\vmax expressionStatement, keep
                        [\vmax postfixExpression, keep 
                          [Identifier\\\token{printf}, keep]
                          [\vmax *, keep
                            [\token{(}, keep] 
                                  [StringLiteral\\\token{"Hello world!\textbackslash{}n"}, keep] 
                            [\token{)}, keep] 
                          ]
                        ]
                      [\token{;}, keep] 
                    ]
                  ]
                  [\token{\}}, keep] 
                ]
    ]
  ]
]
\end{forest}
\end{lrbox}

\newsavebox\hwchoistbox
\begin{lrbox}{\hwchoistbox}
\begin{lstlisting}[language=clike]
int main() {
  printf("Hello world!\n");
}
\end{lstlisting}
\end{lrbox}

\begin{figure}[t]
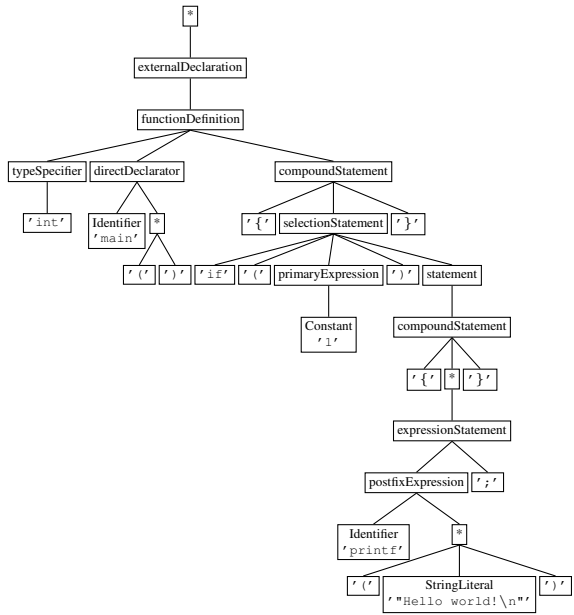

\centering
\subfloat[]{\usebox\hwcbox\label{fig:exampleC}}
\\
\subfloat[]{\usebox\hwctreebox\label{fig:exampleCParseTree}}
\caption{An overly complicated ``Hello World'' program:
\protect\subref*{fig:exampleC}~written in C and
\protect\subref*{fig:exampleCParseTree}~its parse tree.}
\label{fig:hwc}
\end{figure}

\begin{figure}[t]
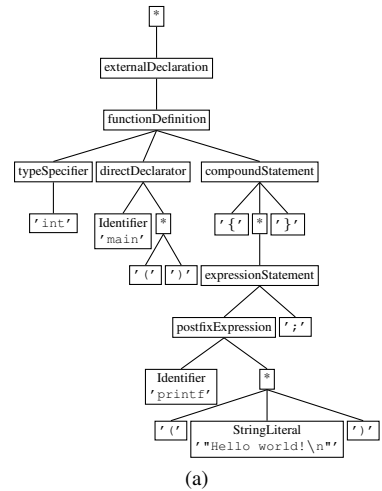

\centering
\subfloat[]{\usebox\hwchoisttreebox\label{fig:exampleCHoistParseTree}}
\\
\subfloat[]{\usebox\hwchoistbox\label{fig:exampleCHoist}}
\caption{The example program of Figure~\ref{fig:hwc} minimized to keep printing the ``Hello world!'' message:
\protect\subref*{fig:exampleCHoistParseTree}~the parse tree with hoisting applied and
\protect\subref*{fig:exampleCHoist}~the C program serialized from the tree.}
\label{fig:hwc-hoist}
\end{figure}

A simple example of this suboptimal structure-preserving behavior is shown in Figure~\ref{fig:hwc}.
The C program in Figure~\ref{fig:exampleC} prints the classic ``Hello world!'' message, but it wraps the printing in an \emph{if} statement where the predicate always evaluates to true.
If we take this program as a test case and define the printing of the ``Hello world!'' message as interesting, then we can try and minimize it\footnote{This is an example where the interesting property of the test case is \emph{not} a program failure.}.
Figure~\ref{fig:exampleCParseTree} shows the parse tree of the program, generated by a parser using a context-free grammar of the C programming language and preprocessed for compactness (most notably, squeezing and recursion flattening have been applied).
Unfortunately, {\HDD} cannot reduce this test case any further (especially if replacement with minimal syntactically correct fragment is used when pruning subtrees) as removing any of the nodes would either yield a syntactically incorrect test case, or one that does not print the message, making it uninteresting\footnote{{\DDMIN} (either line or character-based) would not be able to remove the unnecessary \emph{if} from around the printing either.}.

Theoretically, both {\HDD} and the underlying {\DDMIN} algorithms could be modified to give $n$-tree-minimal results, but that would lead to exponential complexity, which is impractical.
Thus, we propose another approach called \emph{hoisting}.

We can observe that there are recurring structures in the parse tree, subtrees rooted at nodes with identical labels, denoting the derivation of the same non-terminal of the grammar.
The assumption of hoisting is that one such subtree may be replaced with another without losing syntactic correctness, and that subtrees whose roots are in ancestor-descendant relationship may be good candidates for reduction.
Of course, the testing function has to confirm (or reject) whether such a transformation keeps the resulting test case interesting.

In the tree of Figure~\ref{fig:exampleCParseTree}, there is one pair of such subtrees, those rooted at nodes labeled as \emph{compoundStatement}.
Figure~\ref{fig:exampleCHoistParseTree} shows a transformed tree where the descendant subtree is hoisted to replace all the structures that enclosed it.
When this tree is serialized into the form of a C program (see Figure~\ref{fig:exampleCHoist}), it becomes apparent that, in this case, this transformation was indeed useful and we got a smaller and still interesting test case.

\newsavebox\pijbox
\begin{lrbox}{\pijbox}
\begin{lstlisting}[language=javalike]
public class LocalizedPi {
  private static String decSep(String locale) {
    if (locale.equals("en")) {
      return ".";
    }
    throw new Exception("Unsupported locale");
  }
  private static String formatParts(String intPart,
    String fracPart, String decSep) {
    return intPart.concat(decSep).concat(fracPart);
  }
  public static void main(String[] args) {
    String pi = formatParts(
      "3", "14", decSep(args[0]));
    System.out.println(pi);
  }
}
\end{lstlisting}
\end{lrbox}

\newsavebox\pijhddbox
\begin{lrbox}{\pijhddbox}
\begin{lstlisting}[language=javalike]
class LocalizedPi {
  static String decSep(String a) {
    throw new Exception("Unsupported locale");
  }
  static String formatParts(String intPart,
    String fracPart, String decSep) {
    return intPart.concat(decSep).concat(fracPart);
  }
  public static void main(String[] args) {
    String a = formatParts("", "", decSep(args[0]));
  }
}
\end{lstlisting}
\end{lrbox}

\newsavebox\pijhoistbox
\begin{lrbox}{\pijhoistbox}
\begin{lstlisting}[language=javalike]
class LocalizedPi {
  static String decSep(String a) {
    throw new Exception("Unsupported locale");
  }
  public static void main(String[] args) {
    String a = decSep(args[0]);
  }
}
\end{lstlisting}
\end{lrbox}

\begin{figure}[t]
\centering
\subfloat[]{\hspace{0.28cm}\scalebox{0.9}{\usebox\pijbox\label{fig:exampleJava}}}
\\
\subfloat[]{\hspace{0.42cm}\scalebox{0.9}{\usebox\pijhddbox\label{fig:exampleJavaHDD}}}
\\
\subfloat[]{\hspace{-0.5cm}\scalebox{0.9}{\usebox\pijhoistbox\label{fig:exampleJavaHoist}}}
\\
\caption{A program to print the rounded value of $\pi$ in a locale-specific format:
\protect\subref*{fig:exampleJava}~written in Java,
\protect\subref*{fig:exampleJavaHDD}~minimized with {\HDD} to keep the program throw an uncaught exception if an unsupported locale is specified on the command line, and
\protect\subref*{fig:exampleJavaHoist}~minimized with hoisting applied before \HDD.}
\label{fig:pij}
\end{figure}

When discussing the idea of hoisting with fellow researchers, the argument was often raised that such a transformation is only good for removing some minor syntactic elements from the result, like a dangling semicolon or a pair of superfluous braces, etc.
The example of Figures~\ref{fig:hwc} and~\ref{fig:hwc-hoist} does not seem to contradict such arguments.
However, Figure~\ref{fig:exampleJava} shows another example, a program written in Java, that prints the rounded value of $\pi$ in a localized format, provided that the specified locale is supported, and throws an exception otherwise.

As presented, the program only supports the \emph{en} locale.
We shall take this program as a test case and the testing function shall check whether the program throws an exception when invoked with an unsupported locale (e.g., \emph{hu}).
If {\HDD} is used to minimize this test case, it will be able to remove some parts of the program, but most of the original structure will remain in the output.
(Because the exception that needs to be thrown is in the \emph{decSep} method, which is called inside a call to the \emph{formatParts} method, both methods are forced to be kept in the reduced test case.)
The parse tree for this program would be too big to be presented as an example, so we only show the \HDD-reduced Java program in Figure~\ref{fig:exampleJavaHDD}, which also displays precisely what {\HDD} can and what it cannot prune away.
However, if hoisting is used \emph{before} \HDD, it can pave the way for the latter reduction technique by hoisting the call to \emph{decSep} to replace the enclosing call to \emph{formatParts}, thus allowing the complete removal of the definition of \emph{formatParts}.
(In this example, the method calls are the recurring structures that are in ancestor-descendant relation in the tree.)
The result of the combined application of the two techniques is presented in Figure~\ref{fig:exampleJavaHoist}.

Note that \emph{formatParts} could be constructed arbitrarily complex, making the theoretical potential of hoisting considerably higher than the removal of the anecdotical semicolons or curly braces.
Also note, however, that hoisting cannot achieve this alone, as it ``only'' moves subtrees higher up the tree, but it has to cooperate with \HDD.

\subsection{Transformation-based Minimization}
\label{sec:hoisting-trmin}

\begin{figure*}[th!]
\small
\makebox[\linewidth]{\hrulefill}
Let $D$ denote the set of all potential test case elements, and let $\delta \in D$ denote one element of that set.\\
Let $test$ and $c_\fail = \{\delta_1, \dots, \delta_n\} \subseteq D$, and $test(c_\fail) = \fail$ holds.\\
Let $\tau$ and $\|\cdot\|$ be given such that $\forall \delta \in D \cdot \forall \delta' \in \tau(\delta) \cdot \|\delta'\| < \|\delta\|$ holds.\\
The goal is to find $t_\fail = tmin^\tau(c_\fail)$ such that $test(\bar{t}_\fail(c_\fail)) = \fail$ and $t_\fail$ is 1-maximal.\\
The \textit{transformation-based minimizing algorithm} $tmin^\tau(c)$ is
\begin{align*}
tmin^\tau(c_\fail) &= tmin_2^\tau(c_\fail, id_D)  \text{ where} \\
tmin_2^\tau(c_\fail, t'_\fail) &= \left\{ \begin{alignedat}{-1}
  & tmin_2^\tau(c_\fail, t'_\fail[\delta \mapsto \delta']) && \text{if } \exists \delta \in c_\fail \cdot \exists \delta' \in \tau(t'_\fail(\delta)) \cdot test(\bar{t'_\fail}[\delta \mapsto \delta'](c_\fail)) = \fail \\
  & t'_\fail                                               && \text{otherwise.}
\end{alignedat} \right.
\end{align*}
The recursion invariant (and thus precondition) for $tmin_2^\tau$ is $test(\bar{t'_\fail}(c_\fail)) = \fail$.\\
\makebox[\linewidth]{\hrulefill}
\caption{The Transformation-based Minimizing Algorithm.}
\label{fig:tmin}
\end{figure*}

To formalize the ideas described and motivated above, we build on and extend the notations and terminology of delta debugging (as given in Section~\ref{sec:background}) to introduce transformation-based minimization first.

In the context of delta debugging, a test case is always composed of a subset of the elements of the initial failing configuration.
The testing function is also defined for the subsets of $c_\fail$ only.
However, it is useful (actually, necessary) if we can also determine the outcome of a program run on a set of elements, even if some of them are not part of the initial configuration (e.g., in the case of hoisting, when an element -- a node -- is replaced by another element -- another node further down the hierarchy --, which is part of the tree, but is not a member of the initial set).
Therefore, we extend the definitions of delta debugging~\cite{zeller-tse-2002} as follows.

Let $D$ denote the set of all potential test case elements, and let $\delta \in D$ denote one element of that set, i.e., a test case element.
A test case or configuration is denoted as $c \subseteq D$.
A testing function $test : 2^D \rightarrow \{\fail, \pass, \unres\}$ shall determine for any test case whether it produces the failure in question.
The initial failing configuration is denoted as $c_\fail = \{\delta_1, \dots, \delta_n\} \subseteq D$, and $test(c_\fail) = \fail$ holds.

As $c_\fail$ is a subset of a potentially larger set $D$, we allow for \emph{transformations} that can not only remove, but also \emph{replace} elements in the configuration.
We use the following definitions and notations for transformations:

A function $t : D \rightarrow D$ is a transformation of test case elements, and the identity transformation of test case elements is $id_D : D \rightarrow D; \delta \mapsto \delta$.
We also define the application of a test case element transformation to test cases (or configurations) as $\bar{t} : 2^D \rightarrow 2^D; c \mapsto \{t(\delta) : \delta \in c\}$ (e.g., $\overline{id}_D(c_\fail) = c_\fail$).
And a transformation that is derived from another transformation by changing the mapping of one test case element is defined as
$$t[\delta' \mapsto \delta''] : D \rightarrow D; \delta \mapsto \left\{ \begin{alignedat}{-1}
  & \delta''    && \textit{if } \delta = \delta' \\
  & t(\delta)   && \textit{otherwise.}
\end{alignedat} \right.$$

In the examples of Section~\ref{sec:hoisting-ex}, the transformations that could be applied were quite straightforward.
There was only one \emph{compoundStatement} and one method call that could potentially replace their parents.
In a general case, however, a test case element may have multiple replacement candidates (or none at all).
This is formalized using a function $\tau : D \rightarrow 2^D$ that maps test case elements to their transformed candidates.

Finally, as test cases are not necessarily the subsets of the initial failing configuration, minimality cannot be defined in terms of the subset relation anymore.
Thus, we expect a $\|\cdot\|$ measure to exist on set $D$.

If all transformation candidates in $\tau$ are potentially reducing the size of a configuration according to the measure $\|\cdot\|$, i.e., $\forall \delta \in D \cdot \forall \delta' \in \tau(\delta) \cdot \|\delta'\| < \|\delta\|$ holds, then in order to minimize the test case we have to maximize the replacements applied to the elements of the initial configuration (even transitively) while ensuring that the so-transformed test case remains interesting.
Just like it is true for {\DDMIN} that searching for the global optimum is impractical, so is it also true for transformation-based minimization.
Therefore, our actual goal is to find a local optimum, a \emph{1-maximal} transformation $t_\fail$ such that $\forall \delta \in c_\fail \cdot \forall \delta' \in \tau(t_\fail(\delta)) \cdot test(\bar{t}_\fail[\delta \mapsto \delta'](c_\fail)) \neq \fail $ holds.

Figure~\ref{fig:tmin} wraps up this subsection and formalizes the transformation-based minimizing algorithm \TMIN$^\tau$, worded in the likeness of \DDMIN.

\subsection{Hoisting and HDD}
\label{sec:hoisting-and-hdd}

The transformation-based minimizing algorithm gives us a framework to formulate hoisting.
The key to this is to define hoisting as a transformation of tree nodes.
More precisely, to define those nodes in the tree representation of the input that can act as replacement candidates for their ancestors.
The formula in Figure~\ref{fig:hoistables}, $\chi(n)$, is one possible way to define these candidates, i.e., the hoistable descendants of a node $n$.
$\chi(n)$ is given in terms of two auxiliary functions, of which \alg{children}($n$) is trivial, giving the direct descendants of a node, whereas \alg{compatible}($n$, $n'$) leaves some space for interpretation.
In an extreme case, any two nodes could be considered compatible, but that would rarely be useful.
If the tree representation of the input is built using a context-free grammar, as motivated in Section~\ref{sec:hoisting-ex}, then a natural interpretation is to regard identically-labeled nodes (i.e., subtrees of derivations of the same non-terminal of the grammar) as compatible.

\begin{figure}[t]
\centering
\small
\begin{align*}
\chi(n) &= \bigcup_{n' \in children(n)} \chi'(n,n') \\
\chi'(n,n') &= \left\{ \begin{alignedat}{-1}
  & \{n'\}                                      && \text{if } compatible(n,n') \\
  & \bigcup_{n'' \in children(n')} \chi'(n,n'') && \text{otherwise}
\end{alignedat} \right.
\end{align*}
\caption{$\chi(n)$, the potentially hoistable descendants of node $n$.}
\label{fig:hoistables}
\end{figure}

\begin{figure}[t]
\begin{lstlisting}[language=pseudo]
procedure $\HOIST(\textit{input\_tree})$
    $\textit{level} \leftarrow 0$
    $\textit{nodes} \leftarrow \TAGNODES(\textit{input\_tree}, \textit{level})$
    while $\textit{nodes} \neq \emptyset$ do
        $\textit{hoisting} \leftarrow \TMIN^\chi(\textit{nodes})$
        $\TRANSFORM(\textit{input\_tree}, \textit{level}, \textit{hoisting})$
        $\textit{level} \leftarrow \textit{level} + 1$
        $\textit{nodes} \leftarrow \TAGNODES(\textit{input\_tree}, \textit{level})$
    end while
end procedure
\end{lstlisting}
\caption{The Hoisting Algorithm.}
\label{fig:hoist}
\end{figure}

The natural measure to use for tree nodes is based on the size of subtrees, i.e., the number assigned by the measure to a node $n$ equals the number of nodes in the subtree of $n$.
It is obvious that all transformation candidates returned by $\chi(n)$ reduce the size of the configuration according to this measure, as expected by the definition of \TMIN.

Now, with the help of \TMIN$^\chi$, we can introduce a hierarchical algorithm, called \HOIST, that works its way through the tree from the root to the leaves, and uses \TMIN$^\chi$ to find the hoisting transformations at each level.
Candidates found by \TMIN$^\chi$ are prioritized by their distance to the ancestor, further nodes give higher priority.
The pseudocode of the algorithm is presented in Figure~\ref{fig:hoist}.
The structure of {\HOIST} is similar to that of \HDD:
both contain a loop through the levels of the tree, and inside the loop, both perform a minimization step (\TMIN$^\chi$ vs. \DDMIN) and the application of its result to the tree (via the {\TRANSFORM} and {\PRUNE} auxiliary functions).

As discussed at the example of Figure~\ref{fig:pij}, although {\HOIST} can achieve reduction on its own, it is expected to work best if used in combination with \HDD, e.g., by using {\HOIST} as a preprocessing step.
However, inspired by the similarities between the two algorithms, we can think of other ways of combination as well.
E.g., the bodies of the loops of the two algorithms can be interlaced, performing both the {\DDMIN} and \TMIN$^\chi$-based minimizations at each level.
One way to formulate this idea is shown in Figure~\ref{fig:hddh}, in the algorithm named \HDDH.

\begin{figure}[t]
\begin{lstlisting}[language=pseudo]
procedure $\HDDH(\textit{input\_tree})$
    $\textit{level} \leftarrow 0$
    $\textit{nodes} \leftarrow \TAGNODES(\textit{input\_tree}, \textit{level})$
    while $\textit{nodes} \neq \emptyset$ do
        $\textit{minconfig} \leftarrow \DDMIN(\textit{nodes})$
        $\PRUNE(\textit{input\_tree}, \textit{level}, \textit{minconfig})$
        $\textit{hoisting} \leftarrow \TMIN^\chi(\textit{minconfig})$
        $\TRANSFORM(\textit{input\_tree}, \textit{level}, \textit{hoisting})$
        $\textit{level} \leftarrow \textit{level} + 1$
        $\textit{nodes} \leftarrow \TAGNODES(\textit{input\_tree}, \textit{level})$
    end while
end procedure
\end{lstlisting}
\caption{The Hierarchical Delta Debugging \& Hoisting Algorithm.}
\label{fig:hddh}
\end{figure}

\section{Experimental Results}
\label{sec:results}

To evaluate the effects of hoisting, we have implemented the above introduced {\HOIST} and {\HDDH} algorithms and published them in the 21.3 release of the Picireny project\footnote{\url{https://github.com/renatahodovan/picireny}}.
Picireny is a hierarchical test case reduction framework written in Python that supports ANTLR~v4\footnote{\url{https://github.com/antlr/antlr4}} grammars, and already contains an implementation of the {\HDD} algorithm.
In this implementation of hoisting, we consider nodes labeled with the same non-terminal of the grammar as compatible, as discussed in Section~\ref{sec:hoisting-and-hdd}.

We have used the implemented algorithms in four different combinations during our experiments:

\begin{itemize}
\item \emph{\HDD$^*$: hierarchical delta debugging without any hoisting, acted as the baseline.}
\item \emph{\HOIST$^*$+\HDD$^*$: hoisting was applied as a preprocessing step to \HDD$^*$.}
\item \emph{\HDDH$^*$: hoisting interlaced with \HDD$^*$.}
\item \emph{\HOIST$^*$+\HDDH$^*$: \HOIST$^*$ and \HDDH$^*$ algorithms are not mutually exclusive, thus we have used them in sequence.}
\end{itemize}

In all cases, the asterisk superscript denotes the fixed-point iteration of the marked algorithm.

As inputs, we have collected test cases from different sources.
The first (small) set of test cases is composed of the examples from Section~\ref{sec:hoisting-ex} (\emph{helloworld.c} and \emph{LocalizedPi.java}, with properties to keep as described above).
The second set of test cases have already been used in the literature for benchmarking reduction: the Perses Test Suite\footnote{\url{https://github.com/perses-project/perses}}{\saveFN\persesFN} contains fuzzer-generated C sources that cause various internal compiler errors in the Clang and GCC compilers\footnote{The Perses Test Suite comes with a docker environment provided by its authors. The environment is presumed to contain all compiler versions and tools required to reproduce the issue of each test case in the suite. However, that turned out not to be the case in practice. Thus, we have only used those test cases for evaluation that worked as expected at the time of writing this paper.}.
Finally, as a third set, we have composed a new suite of JavaScript sources, also generated with fuzzing, that cause failures in the JerryScript lightweight JavaScript engine (which will be referred to as the JerryScript Reduction Test Suite)\footnote{\url{https://github.com/vincedani/jrts}}.
In the case of the latter two test suites, the interesting property of the test cases to keep during reduction is the failure they induce.

For each test case, we have built its parse tree representation using the grammar available for the input format from the official ANTLR~v4 grammars repository\footnote{\url{https://github.com/antlr/grammars-v4}}.
Before evaluating any of the reduction algorithm combinations, we have applied squeezing of linear tree components~\cite{hodovan2017tree} and flattening of recursive structures to the trees~\cite{hodovan2017coarse}.
The experiments were executed on a workstation equipped with an Intel Core i5-9400 CPU clocked at 2.9GHz and 16GB RAM.
The machine was running Ubuntu 20.04 with Linux kernel 5.4.0.

\begin{table*}[p]
\footnotesize
\caption{Examples: Input and Output Sizes (Number of Non-whitespace Characters)}
\label{tab:exampleTestSizes}
\begin{center}
\begin{tabular}{l@{\hspace{.9cm}}r@{\hspace{.9cm}}r@{\hspace{.9cm}}r@{~~}r@{\hspace{.9cm}}r@{~~}r@{\hspace{.9cm}}r@{~~}r}
\toprule
\textbf{Test} & \textbf{Input} & \textbf{\HDD}\boldmath$^*$ & \multicolumn{2}{c@{\hspace{.9cm}}}{\textbf{\HOIST}\boldmath$^*$\textbf{+\HDD}\boldmath$^*$} & \multicolumn{2}{c@{\hspace{.9cm}}}{\textbf{\HDDH}\boldmath$^*$} & \multicolumn{2}{c}{\textbf{\HOIST}\boldmath$^*$\textbf{+\HDDH}\boldmath$^*$} \\
\midrule
helloworld.c     &  42 &  42 &  \textbf{35} & \textit{(-16.67\%)} &  \textbf{35} & \textit{(-16.67\%)} &  \textbf{35} & \textit{(-16.67\%)} \\
LocalizedPi.java & 446 & 359 & \textbf{186} & \textit{(-48.19\%)} & \textbf{186} & \textit{(-48.19\%)} & \textbf{186} & \textit{(-48.19\%)} \\
\bottomrule
\end{tabular}
\end{center}
\end{table*}

\begin{table*}[p]
\footnotesize
\caption{Examples: Number of Test Executions}
\label{tab:exampleStepCounts}
\begin{center}
\begin{tabular}{l@{\hspace{.9cm}}r@{\hspace{.9cm}}r@{~~}r@{\hspace{.9cm}}r@{~~}r@{\hspace{.9cm}}r@{~~}r}
\toprule
\textbf{Test}  & \textbf{\HDD}\boldmath$^*$ & \multicolumn{2}{c@{\hspace{.9cm}}}{\textbf{\HOIST}\boldmath$^*$\textbf{+\HDD}\boldmath$^*$} & \multicolumn{2}{c@{\hspace{.9cm}}}{\textbf{\HDDH}\boldmath$^*$} & \multicolumn{2}{c}{\textbf{\HOIST}\boldmath$^*$\textbf{+\HDDH}\boldmath$^*$} \\
\midrule
helloworld.c     & 32  &  \textbf{26} & \textit{(-18.75\%)} &  51 & \textit{(+59.38\%)} & \textbf{26} & \textit{(-18.75\%)} \\
LocalizedPi.java & 638 & \textbf{302} & \textit{(-52.66\%)} & 588 &  \textit{(-7.84\%)} &         304 & \textit{(-52.35\%)} \\
\bottomrule
\end{tabular}
\end{center}
\end{table*}

\begin{table*}[p]
\footnotesize
\caption{Perses Test Suite: Input and Output Sizes (Number of Non-whitespace Characters)}
\label{tab:hoist}
\begin{center}
\begin{tabular}{l@{\hspace{.9cm}}r@{\hspace{.9cm}}r@{\hspace{.9cm}}r@{~~}r@{\hspace{.9cm}}r@{~~}r@{\hspace{.9cm}}r@{~~}r@{\hspace{.9cm}}r@{\hspace{.9cm}}r}
\toprule
\textbf{Test} & \textbf{Input} & \textbf{\HDD}\boldmath$^*$ & \multicolumn{2}{c@{\hspace{.9cm}}}{\textbf{\HOIST}\boldmath$^*$\textbf{+\HDD}\boldmath$^*$} & \multicolumn{2}{c@{\hspace{.9cm}}}{\textbf{\HDDH}\boldmath$^*$} & \multicolumn{2}{c@{\hspace{.9cm}}}{\textbf{\HOIST}\boldmath$^*$\textbf{+\HDDH}\boldmath$^*$} & \textbf{Perses} & \textbf{Pardis} \\
\midrule
clang-22382 &  65,786 &   582 &         489  & \textit{(-15.98\%)} & \textbf{475} & \textit{(-18.38\%)} &           489  & \textit{(-15.98\%)} &          509 &  1,027 \\
clang-22704 & 597,827 &   168 &         164  &  \textit{(-2.38\%)} &          165 &  \textit{(-1.79\%)} &   \textbf{161} &  \textit{(-4.17\%)} &          246 &    724 \\
clang-23309 & 118,178 & 3,582 &       1,486  & \textit{(-58.51\%)} &        1,677 & \textit{(-53.18\%)} & \textbf{1,416} & \textit{(-60.47\%)} &        1,943 &  5,188 \\
clang-23353 &  94,734 &   374 &         354  &  \textit{(-5.35\%)} &          592 & \textit{(+58.29\%)} &           351  &  \textit{(-6.15\%)} & \textbf{331} &    397 \\
clang-25900 & 245,065 & 1,562 &         986  & \textit{(-36.88\%)} & \textbf{885} & \textit{(-43.34\%)} &           888  & \textit{(-43.15\%)} &          943 &  1,970 \\
clang-26350 & 378,160 & 1,613 &         778  & \textit{(-51.77\%)} & \textbf{585} & \textit{(-63.73\%)} &           760  & \textit{(-52.88\%)} &        1,220 &  1,777 \\
clang-26760 & 588,548 &   586 &         595  &  \textit{(+1.54\%)} & \textbf{297} & \textit{(-49.32\%)} &           582  &  \textit{(-0.68\%)} &          345 &    825 \\
clang-27747 & 409,083 &   419 &         406  &  \textit{(-3.10\%)} & \textbf{377} & \textit{(-10.02\%)} &           415  &  \textit{(-0.95\%)} &          415 &    765 \\
clang-31259 & 137,161 & 2,174 &         814  & \textit{(-62.56\%)} &          947 & \textit{(-56.44\%)} &   \textbf{796} & \textit{(-63.39\%)} &        1,339 &  1,519 \\
gcc-59903   & 166,754 & 1,726 &       1,432  & \textit{(-17.03\%)} &          620 & \textit{(-64.08\%)} &         1,298  & \textit{(-24.80\%)} & \textbf{487} &  3,301 \\
gcc-60116   & 218,223 & 3,788 &       1,185  & \textit{(-68.72\%)} &        1,152 & \textit{(-69.59\%)} &   \textbf{941} & \textit{(-75.16\%)} &        1,823 &  3,852 \\
gcc-61383   & 110,643 & 1,701 &       1,041  & \textit{(-38.80\%)} & \textbf{844} & \textit{(-50.38\%)} &            874 & \textit{(-48.62\%)} &        1,908 &  2,525 \\
gcc-61917   & 254,742 & 1,764 &         575  & \textit{(-67.40\%)} &          885 & \textit{(-49.83\%)} &           570  & \textit{(-67.69\%)} & \textbf{557} &  2,371 \\
gcc-64990   & 439,587 & 2,844 &         561  & \textit{(-80.27\%)} &        1,282 & \textit{(-54.92\%)} &   \textbf{551} & \textit{(-80.63\%)} &          843 &  2,787 \\
gcc-65383   & 125,221 & 1,027 &         543  & \textit{(-47.13\%)} &          490 & \textit{(-52.29\%)} &   \textbf{441} & \textit{(-57.06\%)} &          474 &  1,281 \\
gcc-66186   & 139,087 & 2,614 &         978  & \textit{(-62.59\%)} & \textbf{977} & \textit{(-62.62\%)} &   \textbf{977} & \textit{(-62.62\%)} &        1,124 &  4,144 \\
gcc-66375   & 191,827 & 2,963 &       1,446  & \textit{(-51.20\%)} &        1,439 & \textit{(-51.43\%)} & \textbf{1,430} & \textit{(-51.74\%)} &        1,594 &  3,918 \\
gcc-70127   & 400,556 &   --- &         992  &        \textit{---} & \textbf{915} &        \textit{---} &            947 &        \textit{---} &          998 &  1,986 \\
gcc-71626   &  14,465 &   168 & \textbf{167} &  \textit{(-0.60\%)} & \textbf{167} &  \textit{(-0.60\%)} &   \textbf{167} &  \textit{(-0.60\%)} & \textbf{167} &    169 \\
\bottomrule
\end{tabular}
\end{center}
\end{table*}

\begin{table*}[p]
\footnotesize
\caption{Perses Test Suite: Number of Test Executions}
\label{tab:hoistStepCount}
\begin{center}
\begin{tabular}{l@{\hspace{.9cm}}r@{\hspace{.9cm}}r@{~~}r@{\hspace{.9cm}}r@{~~}r@{\hspace{.9cm}}r@{~~}r@{\hspace{.9cm}}r@{\hspace{.9cm}}r}
\toprule
\textbf{Test}  & \textbf{\HDD}\boldmath$^*$ & \multicolumn{2}{c@{\hspace{.9cm}}}{\textbf{\HOIST}\boldmath$^*$\textbf{+\HDD}\boldmath$^*$} & \multicolumn{2}{c@{\hspace{.9cm}}}{\textbf{\HDDH}\boldmath$^*$} & \multicolumn{2}{c@{\hspace{.9cm}}}{\textbf{\HOIST}\boldmath$^*$\textbf{+\HDDH}\boldmath$^*$} & \textbf{Perses} & \textbf{Pardis} \\
\midrule
clang-22382 & 14,699 &  9,910 &  \textit{(-32.58\%)} & 12,955 & \textit{(-11.86\%)} &  9,997 &  \textit{(-31.99\%)} &          2,325 & \textbf{2,331} \\
clang-22704 & 10,540 & 21,094 & \textit{(+100.13\%)} & 10,474 &  \textit{(-0.63\%)} & 21,180 & \textit{(+100.95\%)} & \textbf{1,890} &          4,705 \\
clang-23309 & 24,630 & 16,025 &  \textit{(-34.94\%)} & 19,833 & \textit{(-19.48\%)} & 15,828 &  \textit{(-35.74\%)} &          4,147 & \textbf{3,960} \\
clang-23353 & 14,598 & 30,114 & \textit{(+106.29\%)} & 14,662 &  \textit{(+0.44\%)} & 30,182 & \textit{(+106.75\%)} & \textbf{2,288} &          2,324 \\
clang-25900 & 14,766 &  9,983 &  \textit{(-32.39\%)} & 12,510 & \textit{(-15.28\%)} &  9,865 &  \textit{(-33.19\%)} & \textbf{2,115} &          2,431 \\
clang-26350 & 16,789 & 18,851 &   \textit{(12.28\%)} & 14,831 & \textit{(-11.66\%)} & 19,847 &  \textit{(+18.21\%)} & \textbf{3,976} &          8,835 \\
clang-26760 & 12,957 & 11,808 &   \textit{(-8.87\%)} & 11,884 &  \textit{(-8.28\%)} & 11,835 &   \textit{(-8.66\%)} &          1,933 & \textbf{1,828} \\
clang-27747 &  7,174 & 13,899 &  \textit{(+93.74\%)} &  6,601 &  \textit{(-7.99\%)} & 13,911 &  \textit{(+93.91\%)} &          1,559 & \textbf{1,545} \\
clang-31259 & 19,239 &  8,791 &  \textit{(-54.31\%)} & 15,914 & \textit{(-17.28\%)} &  8,992 &  \textit{(-53.26\%)} &          2,230 & \textbf{2,107} \\
gcc-59903   & 18,935 & 12,554 &  \textit{(-33.70\%)} & 18,381 &  \textit{(-2.93\%)} & 12,345 &  \textit{(-34.80\%)} &          3,879 & \textbf{3,761} \\
gcc-60116   & 23,844 & 12,740 &  \textit{(-46.57\%)} & 17,153 & \textit{(-28.06\%)} & 12,041 &  \textit{(-49.50\%)} & \textbf{4,394} &          4,975 \\
gcc-61383   & 17,286 & 11,802 &  \textit{(-31.73\%)} & 15,350 & \textit{(-11.20\%)} & 11,984 &  \textit{(-30.67\%)} & \textbf{3,326} &          4,275 \\
gcc-61917   & 17,455 &  8,432 &  \textit{(-51.69\%)} & 13,769 & \textit{(-21.12\%)} &  8,525 &  \textit{(-51.16\%)} & \textbf{2,802} &          3,566 \\
gcc-64990   & 19,624 & 10,533 &  \textit{(-46.33\%)} & 17,565 & \textit{(-10.49\%)} & 10,548 &  \textit{(-46.25\%)} & \textbf{2,690} &          3,799 \\
gcc-65383   & 16,239 &  6,524 &  \textit{(-59.83\%)} & 11,801 & \textit{(-27.33\%)} &  6,334 &  \textit{(-61.00\%)} & \textbf{2,097} &          2,843 \\
gcc-66186   & 16,181 & 13,771 &  \textit{(-14.89\%)} & 13,930 & \textit{(-13.91\%)} & 13,762 &  \textit{(-14.95\%)} & \textbf{2,333} &          3,693 \\
gcc-66375   & 21,251 & 16,046 &  \textit{(-24.49\%)} & 18,393 & \textit{(-13.45\%)} & 16,131 &  \textit{(-24.09\%)} & \textbf{2,932} &          3,013 \\
gcc-70127   &    --- & 15,699 &         \textit{---} & 18,330 &      \textit{  ---} & 15,974 &         \textit{---} & \textbf{2,517} &          2,527 \\
gcc-71626   &  4,216 &  6,520 & \textit{(+54.65\%)}  &  4,205 &  \textit{(-0.26\%)} &  6,522 &  \textit{(+54.70\%)} &            567 &   \textbf{274} \\
\bottomrule
\end{tabular}
\end{center}
\end{table*}

\begin{table*}[t!]
\footnotesize
\caption{JerryScript Reduction Test Suite: Input and Output Sizes (Number of Non-whitespace Characters)}
\label{tab:jrts}
\begin{center}
\begin{tabular}{l@{\hspace{.9cm}}r@{\hspace{.9cm}}r@{\hspace{.9cm}}r@{~~}r@{\hspace{.9cm}}r@{~~}r@{\hspace{.9cm}}r@{~~}r@{\hspace{.9cm}}r}
\toprule
\textbf{Test} & \textbf{Input} & \textbf{\HDD}\boldmath$^*$ & \multicolumn{2}{c@{\hspace{.9cm}}}{\textbf{\HOIST}\boldmath$^*$\textbf{+\HDD}\boldmath$^*$} & \multicolumn{2}{c@{\hspace{.9cm}}}{\textbf{\HDDH}\boldmath$^*$} & \multicolumn{2}{c@{\hspace{.9cm}}}{\textbf{\HOIST}\boldmath$^*$\textbf{+\HDDH}\boldmath$^*$} & \textbf{Perses} \\
\midrule
jerry-3299 & 1,208 &          92 &  \textbf{89} &  \textit{(-3.26\%)} &  \textbf{89} &  \textit{(-3.26\%)} &  \textbf{89} &  \textit{(-3.26\%)} & 152 \\
jerry-3361 & 1,520 &          97 &  \textbf{95} &  \textit{(-2.06\%)} &  \textbf{95} &  \textit{(-2.06\%)} &  \textbf{95} &  \textit{(-2.06\%)} & 136 \\
jerry-3376 & 4,647 &          70 &  \textbf{35} & \textit{(-50.00\%)} &           37 & \textit{(-47.14\%)} &  \textbf{35} & \textit{(-50.00\%)} & 202 \\
jerry-3408 & 2,100 &          62 &  \textbf{54} & \textit{(-12.90\%)} &  \textbf{54} & \textit{(-12.90\%)} &  \textbf{54} & \textit{(-12.90\%)} & 102 \\
jerry-3431 &   648 &          28 &  \textbf{27} &  \textit{(-3.57\%)} &  \textbf{27} &  \textit{(-3.57\%)} &  \textbf{27} &  \textit{(-3.57\%)} &  62 \\
jerry-3433 &   652 & \textbf{18} &  \textbf{18} &   \textit{(0.00\%)} &  \textbf{18} &   \textit{(0.00\%)} &  \textbf{18} &   \textit{(0.00\%)} &  67 \\
jerry-3437 & 4,623 &          34 &  \textbf{18} & \textit{(-47.06\%)} &  \textbf{18} & \textit{(-47.06\%)} &  \textbf{18} & \textit{(-47.06\%)} &  33 \\
jerry-3479 & 3,998 &          94 &  \textbf{89} &  \textit{(-5.32\%)} &  \textbf{89} &  \textit{(-5.32\%)} &  \textbf{89} &  \textit{(-5.32\%)} & 413 \\
jerry-3483 &   326 & \textbf{38} &  \textbf{38} &   \textit{(0.00\%)} &  \textbf{38} &   \textit{(0.00\%)} &  \textbf{38} &   \textit{(0.00\%)} &  50 \\
jerry-3506 & 2,735 & \textbf{52} &  \textbf{52} &   \textit{(0.00\%)} &  \textbf{52} &   \textit{(0.00\%)} &  \textbf{52} &   \textit{(0.00\%)} & 135 \\
jerry-3523 & 2,802 &          63 &  \textbf{48} & \textit{(-23.81\%)} &  \textbf{48} & \textit{(-23.81\%)} &  \textbf{48} & \textit{(-23.81\%)} & 120 \\
jerry-3534 & 1,409 &          96 &  \textbf{80} & \textit{(-16.67\%)} &  \textbf{80} & \textit{(-16.67\%)} &  \textbf{80} & \textit{(-16.67\%)} & 115 \\
jerry-3536 &   592 &         123 & \textbf{120} &  \textit{(-2.44\%)} & \textbf{120} &  \textit{(-2.44\%)} & \textbf{120} &  \textit{(-2.44\%)} & 135 \\
\bottomrule
\end{tabular}
\end{center}
\end{table*}

\begin{table*}[t!]
\footnotesize
\caption{JerryScript Reduction Test Suite: Number of Test Executions}
\label{tab:jrtsStepCount}
\begin{center}
  \begin{tabular}{l@{\hspace{.9cm}}r@{\hspace{.9cm}}r@{~~}r@{\hspace{.9cm}}r@{~~}r@{\hspace{.9cm}}r@{~~}r@{~~}r}
\toprule
\textbf{Test} & \textbf{\HDD}\boldmath$^*$ & \multicolumn{2}{c@{\hspace{.9cm}}}{\textbf{\HOIST}\boldmath$^*$\textbf{+\HDD}\boldmath$^*$} & \multicolumn{2}{c@{\hspace{.9cm}}}{\textbf{\HDDH}\boldmath$^*$} & \multicolumn{2}{c}{\textbf{\HOIST}\boldmath$^*$\textbf{+\HDDH}\boldmath$^*$} & \textbf{Perses} \\
\midrule
jerry-3299 &          176 & 228 &  \textit{(+29.55\%)} &          192 &  \textit{(+9.09\%)} & 251 &  \textit{(+42.61\%)} & \textbf{169} \\
jerry-3361 & \textbf{144} & 254 &  \textit{(+76.39\%)} &          154 &  \textit{(+6.94\%)} & 266 &  \textit{(+84.72\%)} &          199 \\
jerry-3376 &          119 & 412 & \textit{(+246.22\%)} & \textbf{109} &  \textit{(-8.40\%)} & 422 & \textit{(+254.62\%)} &          266 \\
jerry-3408 &          167 & 278 &  \textit{(+66.47\%)} &          178 &  \textit{(+6.59\%)} & 289 &  \textit{(+73.05\%)} & \textbf{160} \\
jerry-3431 &  \textbf{55} & 185 & \textit{(+236.36\%)} &           70 & \textit{(+27.27\%)} & 192 & \textit{(+249.09\%)} &           92 \\
jerry-3433 &  \textbf{18} &  58 & \textit{(+222.22\%)} &           23 & \textit{(+27.78\%)} &  62 & \textit{(+244.44\%)} &           27 \\
jerry-3437 &           49 &  49 &    \textit{(0.00\%)} &  \textbf{48} &  \textit{(-2.04\%)} &  56 &  \textit{(+14.29\%)} &           78 \\
jerry-3479 &          233 & 576 & \textit{(+147.21\%)} & \textbf{230} &  \textit{(-1.29\%)} & 592 & \textit{(+154.08\%)} &          291 \\
jerry-3483 &           69 &  95 &  \textit{(+37.68\%)} &           71 &  \textit{(+2.90\%)} &  97 &  \textit{(+40.58\%)} &  \textbf{64} \\
jerry-3506 & \textbf{115} & 248 & \textit{(+115.65\%)} &          122 &  \textit{(+6.09\%)} & 251 & \textit{(+118.26\%)} &          338 \\
jerry-3523 &          111 & 416 & \textit{(+274.77\%)} &  \textbf{83} & \textit{(-25.23\%)} & 421 & \textit{(+279.28\%)} &          295 \\
jerry-3534 &          173 & 197 &  \textit{(+13.87\%)} & \textbf{149} & \textit{(-13.87\%)} & 200 &  \textit{(+15.61\%)} &          243 \\
jerry-3536 &          150 & 226 &  \textit{(+50.67\%)} &          182 & \textit{(+21.33\%)} & 251 &  \textit{(+67.33\%)} &  \textbf{93} \\
\bottomrule
\end{tabular}
\end{center}
\end{table*}

Table~\ref{tab:exampleTestSizes} shows the sizes of the example test cases, both before and after reduction.
In all cases, size is expressed as the number of non-whitespace characters to avoid bias from indentation or other formatting differences.
For these example inputs, all reduction approaches that use hoisting (i.e., \HOIST$^*$+\HDD$^*$, \HDDH$^*$, and \HOIST$^*$+\HDDH$^*$) give a smaller output than the baseline \HDD$^*$.
Actually, they all give exactly the same output for each input, reducing the size of \emph{helloworld.c} by 16\% and \emph{LocalizedPi.java} by 48\%, compared to \HDD$^*$.
There is a difference in the performance of the algorithms though -- measured in steps, i.e., how many times the testing function was invoked to determine the outcome of a test case --, as shown in Table~\ref{tab:exampleStepCounts}.
For both examples, \HOIST$^*$+\HDD$^*$ and \HOIST$^*$+\HDDH$^*$ required fewer steps than \HDD$^*$ to minimize the input, and \HDDH$^*$ was also faster on \emph{LocalizedPi.java}.
However, \HDDH$^*$ on \emph{helloworld.c} executed more steps than the baseline.
(In each row of these tables, as well as in the tables to follow, bold numbers highlight the best result(s).)

Tables~\ref{tab:hoist} and~\ref{tab:hoistStepCount} show the results of the algorithms on the Perses Test Suite.
Note that for the reduction of the test cases in this test suite, we have used two additional non-\HDD-based state-of-the-art test case reduction tools to give a more comprehensive evaluation of hoisting.
The two tools are Perses{\useFN\persesFN} (revision 34d4dc4, a Java and Kotlin-based implementation of the algorithm of Sun et al.~\cite{perses}) and Pardis\footnote{\url{https://github.com/golnazgh/PARDIS}} (revision b656c6f, by Gharachorlu and Sumner~\cite{pardis}).
For details on the algorithms and their implementations, the reader is referred to the corresponding papers and tool documentations.

As shown in Table~\ref{tab:hoist}, all hoisting-based algorithm combinations produce a smaller output than the baseline in 17 of 19 cases.
(For one input, \emph{gcc-70127}, \HDD$^*$ ran out of memory, but the variants with hoisting correctly finished the minimization. To avoid a biased interpretation of data, we do not consider the correctly performing variants better than the baseline in this case.)
The reduced test cases can be up to 80.27\%, 70.07\%, and 80.63\% smaller (using \HOIST$^*$+\HDD$^*$, \HDDH$^*$, and \HOIST$^*$+\HDDH$^*$, respectively) than the result of \HDD$^*$.
The average improvement is 37.15\%, 38.54\%, and 39.82\%, respectively.
When comparing them to each other, \HDDH$^*$ gives the smallest result in 8 cases, \HOIST$^*$+\HDDH$^*$ produces the smallest output in 9 cases, while there is also a tie, where \HDDH$^*$ and \HOIST$^*$+\HDDH$^*$ produce exactly the same output.
Furthermore there is another tie when all three approaches find the same (local) minimum.
When Pardis and Perses are also included in the comparison, the results of the hoisting-based minimizations are still good enough.
In 15 of the 19 test cases, at least one of \HDDH$^*$ and \HOIST$^*$+\HDDH$^*$ gives smaller results than Perses, and in 11 cases both of them are better.
(The detailed comparison to Pardis is omitted as Perses gave smaller results than Pardis in all cases of this experiment.)

Regarding performance, Table~\ref{tab:hoistStepCount} shows that hoisting can have a positive effect on the number of overall test case evaluations, but not necessarily.
\HDDH$^*$ performed the minimization of 17 inputs faster than the baseline \HDD$^*$, but in those approaches where hoisting was a preprocessing step (namely, in \HOIST$^*$+\HDD$^*$ and \HOIST$^*$+\HDDH$^*$), this improvement is only visible in 13 cases.
Moreover, not even \HDDH$^*$ could overcome Perses or Pardis speed-wise: these two tools were the fastest to produce output for all test cases of the Perses Test Suite.

The experimental results measured on the JerryScript Reduction Test Suite are shown in Tables~\ref{tab:jrts} and~\ref{tab:jrtsStepCount}.
On these 13 inputs, all algorithms work quite similarly with respect to the output size: there are many cases where some or all approaches give identical results.
Still, in 10 cases, all hoisting-based approaches give strictly smaller output than the baseline (by 50\%, 47.14\%, and 50\% in the best case of \HOIST$^*$+\HDD$^*$, \HDDH$^*$, and \HOIST$^*$+\HDDH$^*$, respectively), while in none of the other cases do they give worse results.
On this test suite, the average improvement of the approaches over \HDD$^*$ is 12.85\%, 12.63\%, and 12.85\%, respectively.
Performance-wise, however, this test suite gives significantly different results than the Perses Test Suite.
In the vast majority of the cases, the application of hoisting increased the number of testing steps performed during reduction.
\HOIST$^*$+\HDD$^*$, \HDDH$^*$, and \HOIST$^*$+\HDDH$^*$were slower than \HDD$^*$ in 12, 8, and 13 of the 13 cases, respectively.

We wanted to perform an experiment with Perses and Pardis on this test suite as well.
Although none of these tools have official support for JavaScript inputs at the time of writing this paper, we have managed to extend Perses with the help of the same ANTLR~v4 grammar files as used with Picireny.
(As Perses outperformed Pardis on the previous test suite, we have not spent resources on extending Pardis, eventually.)
The results show that all hoisting-based approaches give strictly smaller outputs than Perses, while producing minimized output faster in 4 of 13 cases.

Based on the experimental data and observations above, we can conclude the following:

\begin{itemize}
\item \emph{On both artificial and real-world inputs, hoisting combined with hierarchical delta debugging gives generally smaller, or at least as small outputs as hierarchical delta debugging alone. Bigger outputs are rare. Minimized test cases can be as small as \sfrac{1}{5} of the output given by traditional hierarchical delta debugging.}
\item \emph{In most of the cases, hierarchical delta debugging interlaced with hoisting (\HDDH$^*$) gives smaller results than non-\HDD-based state-of-the-art reduction techniques.}
\item \emph{The effect of hoisting on the performance of test case reduction is inconclusive. All approaches that use hoisting have shown both improvements and deteriorations in terms of speed in several cases.}
\end{itemize}

\section{Related Work}
\label{sec:relatedwork}

One of the first and most well-known work on automated test case reduction is Delta Debugging by Zeller and Hildebrandt~\cite{zeller-esecfse-1999,hildebrandt-issta-2000,zeller-tse-2002}, minimizing inputs of arbitrary format.
The price of its generality is a potentially lowered performance because of format-breaking incorrect test cases generated and evaluated during the reduction process.
To avoid syntactically broken intermediate test cases, Miserghi and Su proposed to use information about format encoded in context-free grammars, i.e., to convert test cases into a tree representation~\cite{misherghi2006hdd} and apply delta debugging to the levels of the tree.
This Hierarchical Delta Debugging approach helped to remove parts of the test case that aligned with syntactic unit boundaries.
As a further improvement, Miserghi proposed the concept of syntactically correct replacement for nodes of the tree representation that cannot be completely removed from the test case without causing syntax errors~\cite{misherghi2007hdd}.

The formalization of {\HDD} does not detail how to build the tree representation, but its first implementation used traditional context-free grammars to parse the input.
To improve on this, Hodov\'an et al. suggested to use extended context-free grammars for building the tree~\cite{hodovan2016modernizing}, thus creating more balanced representations, which could lead to smaller results and improved performance.
They have also described various tree transformations with the same goal~\cite{hodovan2017coarse,hodovan2017tree}.

Tree-based test case reduction does not necessarily mean subtree removal.
Bruno suggested to use hoisting as an alternative transformation in his framework called SIMP~\cite{bruno2010simp}, which was specifically designed to reduce database-related inputs.
In every reduction step, SIMP tried to replace a node with a compatible descendant.
In a follow-up work that introduced the tool FlexMin, Morton and Bruno extended SIMP with Delta Debugging~\cite{flexmin}.
The main algorithm was the hoisting, while {\DDMIN} was applied only to repeated structures, like lists (column names) and data (string literals).
Instead of manually classifying the nodes into two parts, our algorithm tries to hoist every node if it has at least one compatible descendant.

Sun et al. combined the above approaches into their Perses framework~\cite{perses}.
In their work, they utilized quantifiers and normalized the parse tree producing grammars by rewriting recursive rules to use quantified expressions.
This transformed grammar form was referred to as Perses Normal Form (PNF).
During the reduction, they applied a worklist algorithm, in which non-terminals with more tokens were prioritized over nodes with less token descendants.
In every step, a node was popped from the worklist and reduced according to its type: quantified nodes were reduced with {\DDMIN} while hoisting was applied on non-quantified, regular ones.
They also mentioned that the number of compatible nodes for hoisting can be enormous.
Instead of collecting all candidates and trying to hoist the descendant that has the longest distance from its ancestor, they limited the search space with two constrains: (1) the number of nodes between ancestor-descendants is limited (4 was used in their evaluation); (2) if a compatible descendant has been found, the searching was not continued further down in the hierarchy.

Built upon the ideas introduced in Perses, Gharachorlu and Sumner~\cite{pardis} extended it in a new framework, named Pardis, with an improved queue prioritization algorithm.
This algorithm only considered nullable, i.e., completely removable tree nodes and assigned weights based on a node own token weight instead of its parents.
The other key difference was that eventually they eliminated the hoisting step, since they found it too expensive from performance perspective.

Herfert et al.~\cite{herfertReducer} also combined subtree removal and hoisting in their Generalized Tree Reduction (GTR) algorithm but instead of analyzing a grammar to decide about the applicability of a certain transformation they learned this information from an existing test corpus.
The search-based program repair work of Gazzola et al.~\cite{Gazzola2019AutoRepair} also mentions modifications on the abstract syntax tree, however transformations are given as predefined templates and could be performed randomly without the ancestor-descendant relationship.

The above mentioned works targeted textual inputs, but test case reduction can be applied to other scenarios as well.
Several authors have minimized faulty event sequences originating from various sources:
Scott et al.~\cite{scott2016demi} minimized event sequences of distributed systems,
Clapp et al.~\cite{clapp2016nd3min} aimed at Android GUI event sequences with a variant of \DDMIN.
Moreover, Delta Debugging was even used for the minimization of SMT solver formulas (Brummayer et al.~\cite{brummayer2009smtsolving}).

An interesting analogy between test case reduction and program slicing was recognized by Binkley et al~\cite{binkleyOrbs,binkleyOrbsLimits,yooOrbsPDL,goldTreeOrbs,binkleyTreeVSLineOrbs,binkley2019comparison}.
They have realized that the concepts of slicing (e.g., the program to be sliced or the slicing criterion) can be reformulated as concepts of test case reduction (e.g., the test case or the interestingness property, respectively).
Their approach, called observation-based slicing, avoids the complexities of building a dependency graph representation of a program and can work purely at syntactic level.
Although their approach is not \DDMIN-based, the algorithms show similarities with the ideas of {\DDMIN} and {\HDD}.

We can consider the algorithms as phases in our implementation that process the input via tree traversal.
The pruning phase is identical to {\HDD}, while {\HOIST} phase tries to hoist the furthest compatible descendant from the tree hierarchy.
In addition, other transformations can be implemented easily as new phases into the framework.

\section{Summary}
\label{sec:summary}

In this paper, we have been focusing on the automated reduction of tree-structured test cases.
We have proposed an extension to the well-established pruning-based hierarchical delta debugging technique.
The key idea of the extension, called hoisting, is to allow moving up subtrees in the structure of the test case, thus reducing its size.
We have defined an algorithmic framework for generic transformation-based minimizations and used it to formalize hoisting.
We have also described various ways of how hoisting can be combined with hierarchical delta debugging.

We have prototyped the introduced algorithms and evaluated several combinations of hoisting and pruning on multiple test suites.
The results of our experiments support that hoisting can improve the output of hierarchical delta debugging by producing as much as 80\% smaller minimized tests in the best case.
Moreover, the prototype implementation gave smaller results than other, non-\HDD-based state-of-the-art test case reduction tools in most of the experiments.

As for future work, we have plans to continue this topic of research in various ways.
We wish to conduct further experiments to ensure that the results generalize to inputs that are larger or differently structured compared to those investigated in this paper.
We plan to investigate the combination of hoisting with other {\HDD} algorithm variants, like Coarse~{\HDD}~\cite{hodovan2017coarse} or {\HDD}r~\cite{kiss2018hddr}.
We also aim at speeding up hoisting-extended {\HDD} while not increasing the output size.
Furthermore, we would like to explore additional minimizing transformations.
Finally, we are also interested in the human understandability aspect of reduced test cases, considering both pruning- and transformation-based reduction methods.

\section*{Acknowledgment}

This research was supported by the EU-supported Hungarian national grant GINOP-2.3.2-15-2016-00037
and by grant NKFIH-1279-2/2020 of the Ministry for Innovation and Technology, Hungary.

\bibliographystyle{IEEEtran}
\bibliography{refs}

\end{document}